\documentclass[twocolumn,showpacs,preprintnumbers,amsmath,amssymb,pra]{revtex4}
\usepackage{graphicx}
\usepackage{dcolumn}
\usepackage{bm}
\usepackage{lmodern}
\usepackage{subfigure}
\usepackage{float}
\usepackage{setspace}
\usepackage[colorlinks=true,  linkcolor=blue, urlcolor=blue, citecolor=blue, plainpages=false, pdfpagelabels, breaklinks]{hyperref}

\graphicspath{{./figuras/}}

\begin{document}


\title{Photoassociative ionization of Na inside a storage ring.}

\author{R. Paiva}
\author{R. Muhammad}
\altaffiliation[Also at ]{Department of Applied Physics,Federal Urdu University of Arts, Science and Technology, Islamabad, Pakistan}
%
\author{R. Shiozaki}%
\author{A. L. de Oliveira}
\altaffiliation[Also at ]{Departamento de F\'{i}sica, Universidade do Estado de Santa Catarina-Joinville, SC, Brasil}
\author{O. Morizot}%
\altaffiliation[Also at ]{PIIM, CNRS-Universit\'{e} de Provence (UMR 6633), centre de Saint-J\'{e}r\^{o}me, 13397 Marseille Cedex 20, France.}
\author{V. S. Bagnato}%
\author{K. M. F. Magalh\~{a}es}%

\affiliation{Instituto de F\'{i}sica de S\~{a}o Carlos, Universidade de S\~{a}o Paulo - SP, Brasil}%


\date{\today}

\begin{abstract}
Motivated by recent interest in low dimensional arrays of atoms, we experimentally investigated the way cold collisional processes are affected by the geometry of the considered atomic sample. More specifically, we studied the case of photoassociative ionization (PAI) both in a storage ring where collision is more unidirectional in character and in a trap with clear undefinition of collision axis. First, creating a ring shaped trap (atomotron) we investigated two-color PAI dependence with intensity and polarization of a probing laser. The intensity dependence of the PAI rate was also measured in a magneto-optical trap presenting equivalent temperature and density conditions. Indeed, the results show that in the ring trap, the value of the PAI rate constant is much lower and does not show evidences of saturation, unlike in the case of the 3D-MOT. Cold atomic collisions in storage ring may represent new possibilities for study.
\end{abstract}

\pacs{33.80.-b,34.50.Cx,32.80.-t}

\maketitle


The field of cold atomic collisions followed the overall development of cold atomic physics in general~\cite{Weiner_1999}. The area gave rise to an exciting collection of scientific opportunities and still provides steam for future achievements. Study of novel cold collision geometries may contribute to a better understanding of reaction dynamics, possibly pushing back the frontiers of quantum chemistry.

One of the many questions that motivated the investigation of cold collisions is how an optical field can modify the outcome of a collisional encounter. A natural environment for observation of such an effect is an  atomic trap, like the magneto-optical trap (MOT) for instance. In such a system there is no preferable  axis of collision and only average effects can be measured. Experiments involving atomic beams decelerated by radiation pressure, yet are alternatives to this situation~\cite{Degraffenreid_2000}. However, for collisions in the sub-millikelvin temperature regime, the atomic beam  density is usually much below the one you can reach in a MOT. Recently though, our research group has advised a new way to study cold collisions, producing a ring shaped trap (atomotron)~\cite{Marcassa_2005}, where atomic encounters preferably  take place along one single direction. This trap actually works close to a continuous atomic beam of cold atoms with the tangential direction as the preferable collisional axis.

Concerning cold collision processes, Photoassociative Ionization (PAI)~\cite{Weiner_1989} probably figures amongst the most important ones. PAI is a two step molecular formation resulting in a ionized dimer. PAI was the first measured collisional process observed between cooled atoms~\cite{Lett_1993} and even though it already has been intensively explored~\cite{Weiner_1999}, open questions still remain, specially  related to the relative reaction rates measured in traps and in atomic beams. Due to technical difficulties, temperatures in both systems were never compatible and a fair comparison was never performed for sodium.

Conventional associative ionization proceeds in two steps: excitation of isolated atoms followed by molecular autoionization when the two atoms approach. In contrast, PAI always starts from ground states because atoms move so slowly that radiative lifetimes become short compared to the collision time. PAI is a process that consists in photo excitation of the incoming scattering flux with subsequent photon excitation to a double excited level autoionizing at short range.

In this experiment, by the first time, we realized measurements of the PAI  rate constant at equivalent temperatures but in geometries where either a 3D aspect or a 1D aspect is predominant. We first used a MOT as a collisional system without favorable collision axis and then our storage ring for cold atoms (atomotron) as an experimental system presenting a single predominant collision axis. Both systems operate at equilibrium temperature such that differences in reaction rates can only be associated to the spatial capability of the atoms to explore the collision. On top of being intrinsically interesting, the outcoming results demonstrate new possibilities to explore ultracold chemistry in geometries that are alternatives to those offered by conventional traps.



The experimental set-up 
was already described elsewhere~\cite{Paiva_2009}, therefore only most relevant features will be mentioned here. 
Hot sodium atoms ($\approx$~500~K) are evaporated in an oven and slowed down on their way to the experiment chamber by a Zeeman atomic decelerator~\cite{Phillips_1982}. These atoms are finally captured in a MOT which slowly loads up to an effective atomic density of $10^9$~atoms/cm$^3$.

\begin{figure}%
\includegraphics[width=\columnwidth]{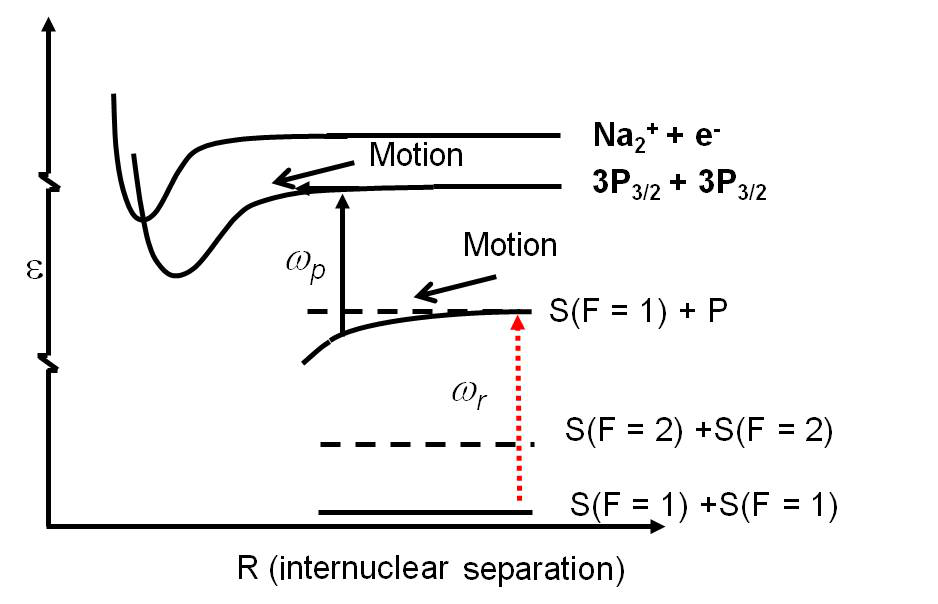}%
\caption{Diagrammatic representation of the photoassociative ionization mechanism in ultracold sodium. In this process two slowly colliding atoms absorb slightly red detuned photons in the trap.}%
\label{fig:diagram}%
\end{figure}

All laser beams are produced by ring dye lasers
. The trapping laser is detuned of $2\Gamma \approx 2\pi \times 10$~MHz to the red of the $F=2-F'=3$ transition (where $\Gamma$ is the natural linewidth of the excited state). Repumping light is then generated by a home-made 1.77~GHz electro-optical modulator 
 producing sidebands in the trapping laser spectrum.
  Typically, the intensity associated to the cooling light is of 8.5~mW/cm$^{2}$ and the resulting intensity of the repumping light of about 3.5~mW/cm$^{2}$. The slowering light 
 is detuned of 60~MHz to the blue of the $F=2-F'=3$ transition. A third dye laser (probe laser) is used in order to perform the second excitation of the two-color PAI process.
The correct unbalance between the cooling laser and the  repumper laser allows us to produce most of the population in the $3 S_{1/2} (F=2)$ or  $3 S_{1/2} (F=1)$ ground state, depending on our choice. We normally work with a strong cooling laser, which provides the starting point for the  collision as  $3 S_{1/2} (F=1) + 3 S_{1/2} (F=1) $ from being further excited by the repumping laser, as described in the diagram of fig.~\ref{fig:diagram}.

The fluorescence of the MOT is collected onto a calibrated photodiode (providing the number of trapped atoms), while a CCD camera makes an image of the trap and provides its dimensions. An ion detector is positioned 50~mm away from the central trapping region in order to collect signal from the PAI.

\begin{figure}%
\includegraphics[width=\columnwidth]{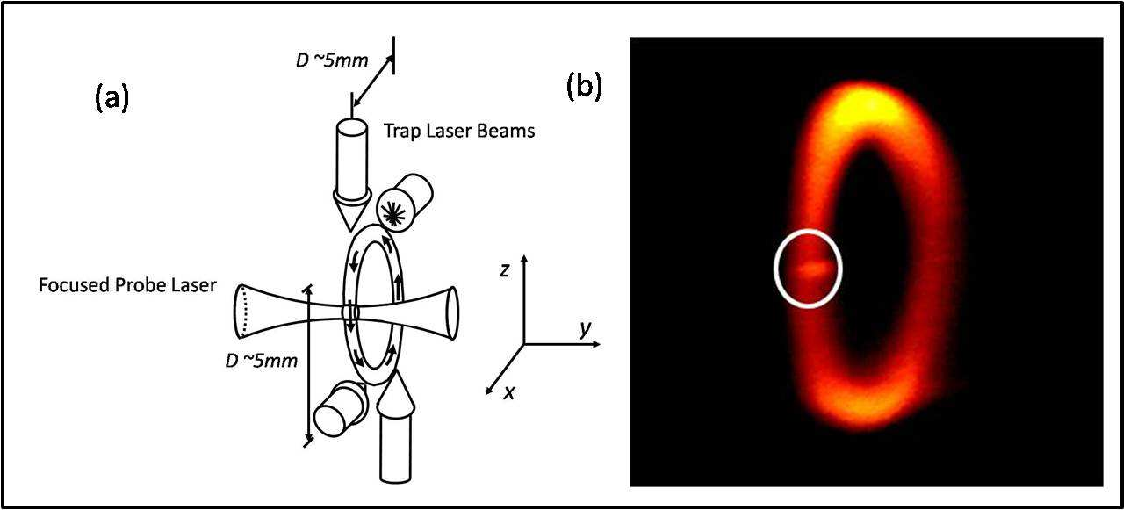} 
\caption{(a) Racetrack misalignment of the Gaussian laser beams to create an atomic distribution in the form of a ring in the xz-plane. (b) An image of the sodium atomotron through a CCD camera. The encircled region is where the probe laser overlaps with the atomotron.}%
\label{fig:atomotron}%
\end{figure}

In some of the experiments we transformed our MOT into an atomotron~\cite{Bagnato_1993,Dias_1996} by inducing small parallel misalignments ($\approx$~5mm) of the x and z beams of the trap, as shown on fig.~\ref{fig:atomotron} (a). A CCD image of the MOT showing trapped atoms under the action of the vortex force is shown on fig.~\ref{fig:atomotron} (b). In that case, the probe laser is intersecting with the atomic ring at right angle and is focused through a long focal length lens in order to have a nearly cylindrical interaction zone. It can deliver up to 200 mW and is focused to a diameter of approximately 100~$\mu$m, restricting the collision volume to be considered.
We performed two-color PAI experiments in two different trap geometries: (i) a usual MOT, where collision axes can be considered as random (3D-case) and (ii) an atomotron presenting a nearly one collision axis (1D-case). In a first set of experiments the PAI rate was measured in the atomotron as a function of the probe polarization, keeping a constant intensity. While in a second set of experiments the PAI rate was measured in both traps, as a function of the probing laser intensity.

PAI is a two step process that has been extensively described in earlier references~\cite{Weiner_1999,Gallagher_1991}. Therefore we will only remind here this process for clarity of the forthcoming discussion, using four stages, fig.~\ref{fig:diagram} can help the visualization of the description: (1) A pair of ultracold atoms in the ground state follows a collisional trajectory in the presence of a radiation field. Note that in our case, due to the relative intensity of cooling and repumping light, most of the trapped atoms are in ground state $3S_{1/2}(F=1)$. (2) One atom of the colliding pair absorbs a photon and is promoted, at long range, to an attractive $C_3/R^3$ potential of an intermediate state ($Na + Na^*$). The two atoms are accelerated towards each other. (3) Absorption of a second photon promotes the pair to a doubly excited state $Na^*+Na^*$. (4) The pair autoionizes to $Na_2^+ + e^-$ at short range.
The whole process can be summed up as :
\begin{equation}
Na + Na + \hbar\omega_1  \rightarrow  Na + Na^* + \hbar\omega_2 \rightarrow  Na_2^* \rightarrow  Na_2^+ + e^-
\label{eq:paiprocess}
\end{equation}

The PAI rate is then given by~\cite{Gallagher_1991} :
\begin{equation}
\frac{d\left[ Na^+_2\right]}{dt} = K_{PAI} \left[Na\right]^2 V,
\label{eq:kpai}
\end{equation}
where $K_{PAI}$ is the cold molecule photoassociative ionization rate constant for a collision happening at temperature T and a fixed probe laser frequency. $\left[Na\right]$ represents the atomic density, and $V$ is the interaction
volume. Note that in the case of the atomotron, the volume is defined by the overlap between the ring and the probe beam. Temperature T of atoms in the ring trap was previously determined as 220$\pm$20~$\mu$K using a time-of-flight technique ~\cite{Araujo_1995} and can be assumed to be practically similar in the case of the MOT.
The tangential mean velocity for the atoms in the ring distribution is approximately 50cm/s. The transversal velocity was already investigated in the past ~\cite{Dias_1996}, and is related to the recoil process during spontaneous emission. We showed that this transverse velocity  distribution is at least five times narrow than the tangential spread ~\cite{Marcassa_2005}. This results in a low transverse velocity component, which gives to the system a 1D character.

In a first set of experiments, the effect of the probe laser polarization on the PAI rate constant was investigated in the atomotron. The probe laser is detuned 625~MHz away to the blue of the $F=2-F'=3$ transition and is continuously shining on the atoms. Its linear polarization is rotated by steps of 10$^\circ$ and the corresponding ion creation rate collected, from which the corresponding PAI rate constant is deduced from eq.~\ref{eq:kpai}.

\begin{figure}%
\includegraphics[width=0.9\columnwidth]{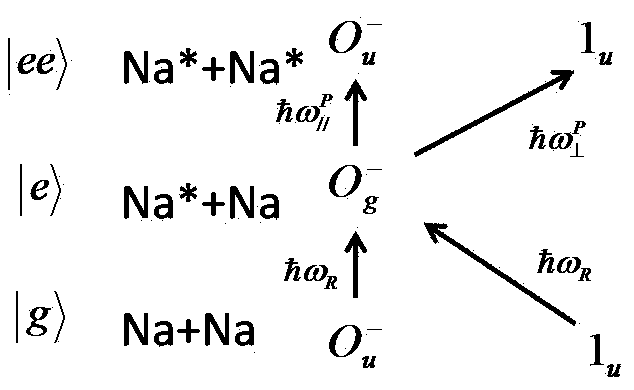} 
\caption{The two-excitation PAI process : First excitation ($g \rightarrow e$) promoted by a circular polarized light ($\omega_{R}$) from the repumper laser. Second excitation ($e \rightarrow ee$) promoted by a linear polarized light from the probe laser, that could be decomposed in  parallel ($i.e.,$ along the tangential direction of the storage ring) ($\omega_{||}^{P}$), which would populate the $O_{u}^{-}$ symmetry state, or perpendicular ($i.e.,$ along its radial direction) ($\omega_{\bot}^{P}$), which would populate the $1_{u}$ symmetry state.}%
\label{fig:polarization}%
\end{figure}

These experiments can be understood in the following way. According to Stwalley et al.~\cite{Stwalley_1978}, within the first 5~GHz detuning of the probe laser, the two-excitation PAI process consists of an initial excitation ($g \rightarrow e$) to a ``pure long-range'' state of $0_g^-$ symmetry. The second excitation ($e\rightarrow ee$) is then promoted by a light with parallel (perpendicular) polarization that changes the quantum number for the total electronic angular momentum projection along the quantization axis by  0 (1). This step promotes the population to the doubly excited state of $0_u^-$ ($1_u$) symmetry. These possibilities of processes can be seen graphically fig. \ref{fig:polarization}. The pair of atoms in the doubly excited state subsequently autoionizes at short range, if the scattering flux reaches the $^3\Sigma_u^+$ molecular state before the occurrence of a spontaneous decay.

In the present experiment, the probe beam, which promotes the second excitation of the PAI process, propagates at a right angle  with respect to the plane of the atomotron. The probe's polarization axis is aligned either parallel ($i.e.,$ along the tangential direction of the storage ring), or perpendicular ($i.e.,$ along its radial direction). Defining the components of the transition dipole moment $\vec{\mu}$  by  $\mu_{||}$ aligned along the molecular internuclear axis and $\mu_{\bot}$ perpendicular to it, one can write the dipole coupling matrix element as $\left\langle e \left|\vec{E}.\vec{\mu}\right|ee\right\rangle$ = $\left\langle e \left|\ E_{||}\mu_{||}\right|ee\right\rangle$ + $\left\langle e \left|\ E_{\bot}\mu_{\bot}\right|ee\right\rangle$.

\begin{figure}%
\includegraphics[width=\columnwidth, height= 5.8cm
]{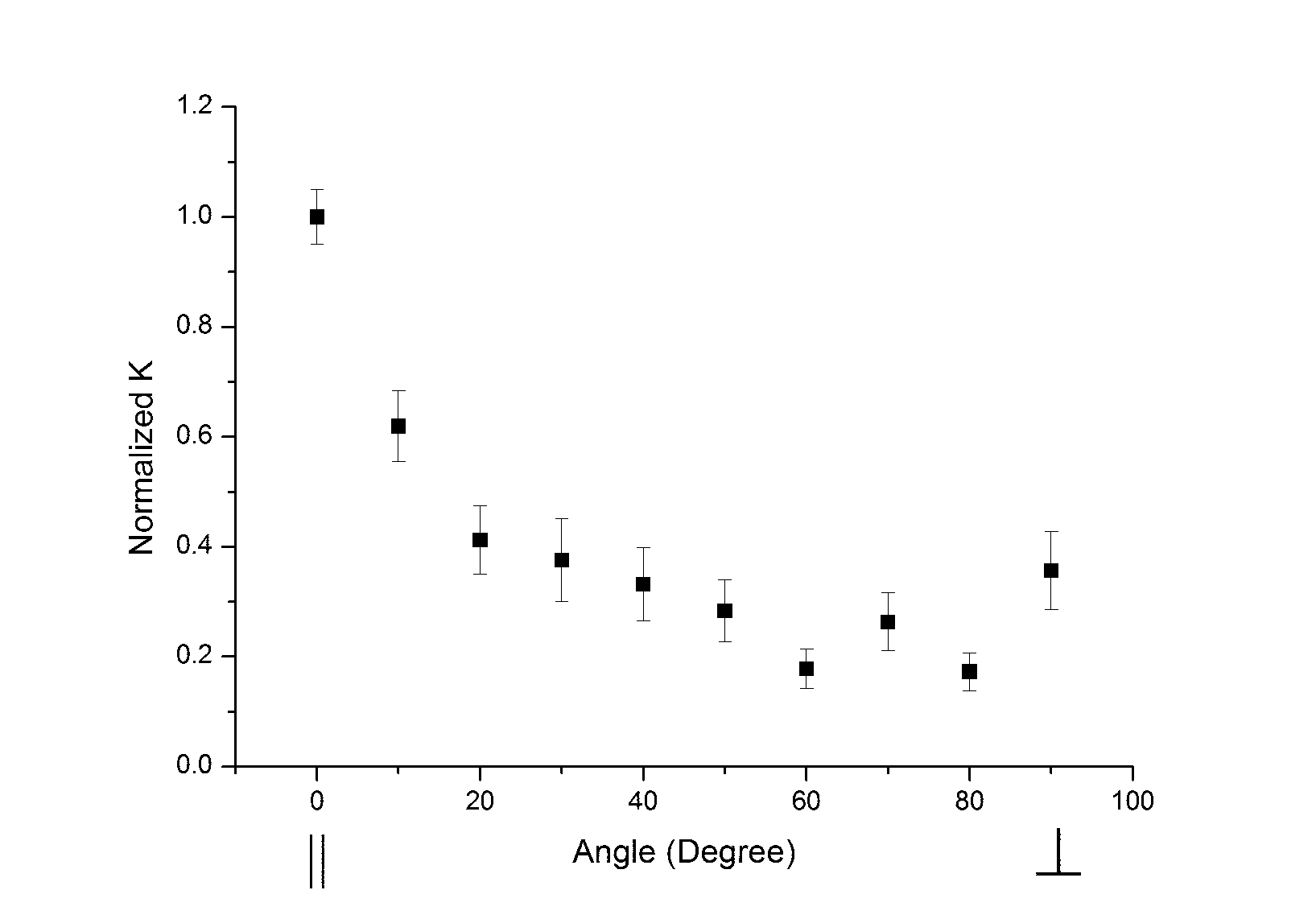}%
\caption{PAI rate constant $K_{PAI}/K_0$ measured in the atomotron as a function of the probe laser polarization angle. Polarization angle of 0$^\circ$ corresponds to light polarization parallel to the preferable direction of the movement and an angle of 90$^\circ$ to a polarization which is orthogonal to the movement direction. Note that for clarity, $K_{PAI}$ is normalized on this graph so that its maximum value corresponds to 1. For the actual value of $K_{PAI}$, refer to fig.~\ref{fig:kinten}. Error bars include standard deviation on ten experimental shots and uncertainty on the density of the cloud.}%
\label{fig:graphpol}%
\end{figure}

In general the effect of the field can then be decomposed along the parallel and perpendicular directions.
As for the parallel direction, the product can be written as
\begin{equation}
E_{||}\mu_{||}= E_z\mu_z
\label{eq:epara}
\end{equation}
since z coincides in our definition to the collision axis of the two molecules. On the other hand the x,y axes are not defined in the experiment (remember that the x,y,z axes are fixed on the collision) as compared with the already defined z-axis . Therefore, the perpendicular component of the polarization is naturally a composition of $ E_x$, $E_y$ for the colliding pair. We therefore  write
\begin{equation}
E_{\bot}\mu_{\bot}= E_x\mu_x + E_y\mu_y
\label{eq:eperpe}
\end{equation}
For an isotropic interacting potential  $\mu_{x} =\mu_{y}  $ and therefore $E_{\bot}\mu_{\bot}= (E_x+ E_y)\mu_x$
When the  polarization of light is parallel, $E_{\bot} =0$ and $ E_{||} =E_0$. That results in a rate constant
\begin{equation}
k_{||}\propto {\left|\left\langle e \left|\stackrel{\rightarrow}{E}. \stackrel{\rightarrow}{\mu}\right|ee\right\rangle\right|}^2 = {E_0}^2{\left|\left\langle e \left|\mu_z\right|ee\right\rangle\right|}^2
\label{eq:kpaipara}
\end{equation}
On the other hand, when the polarization is perpendicular $ E_{||} =0$ and $E_x= E_0 cos\theta $, $E_y= E_0 sin\theta $ (with $\theta$  being the angle between the polarization and x-axis,  resulting in
\begin{equation}
k_{\bot}\propto {\left|\left\langle e \left|\stackrel{\rightarrow}{E}. \stackrel{\rightarrow}{\mu}\right|ee\right\rangle\right|}^2 = \frac{{E_0}^2}{4}{\left|\left\langle e \left|\mu_x\right|ee\right\rangle\right|}^2
\label{eq:kpaiperp}
\end{equation}
where the $\frac{1}{4}$ comes from the average on $\theta$.Therefore, when measuring the rate constants $K_{||}$ and $K_{\bot}$ for identical conditions of probe intensity, one can expect $K_{||} / K_{\bot} =4$, since $K$ is proportional to the squared modulus of the matrix element . And this is what we experimentally observed and represented on fig.~\ref{fig:graphpol}. Although this simple model indeed explains the overall factor of 4, it does not predict the dependence observed for other linear polarizations. A more detailed model would be necessary to fully explain the behavior of $K$ with respect to the polarization angle in intermediate cases, but similar results were obtained in experiments using a decelerated atomic beam of sodium~\cite{RamirezSerrano_2002}.

The polarization experiment therefore certifies that in the atomotrom, the axis of collision is better determined, when compared to a regular trap. This result validates our choice of the atomotron as an ultracold close to unidimensional system, in which PAI rates can finally be reasonably compared to those obtained in a standard 3D MOT.

\begin{figure}%
\includegraphics[width=\columnwidth, height= 5.8cm
]{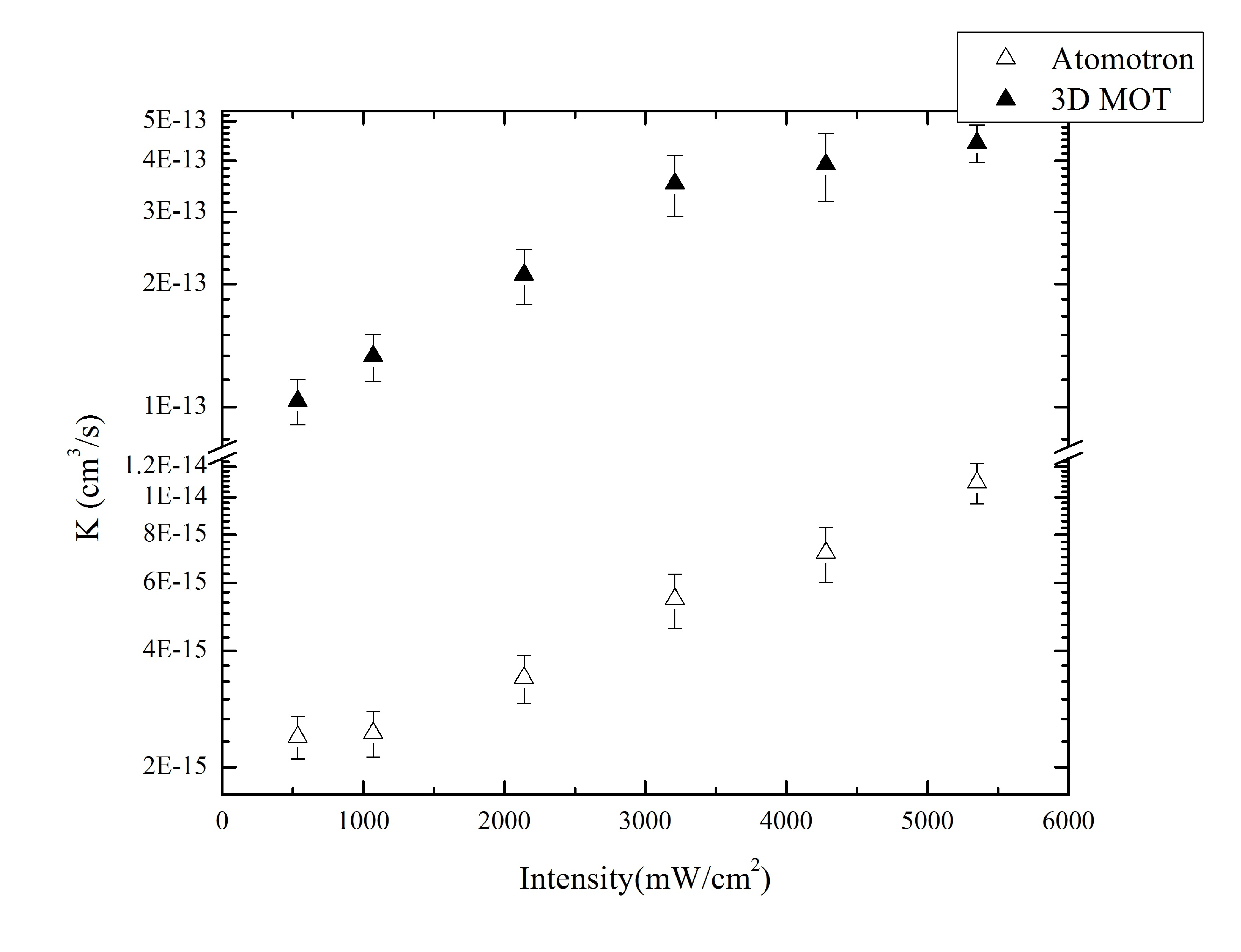}%
\caption{PAI rate constant $K_{PAI}$ as a function of the probe laser intensity. Comparison data between a 3D-MOT (full triangles) and a quasi-1D system (empty triangles). The error bars include standard deviation on ten experimental shots and uncertainty on the density of the cloud. One can observe the difference in magnitude of $K_{PAI}^{3D}$ and $K_{PAI}^{1D}$ (a factor 50 in favor of the 3D-MOT), as well as the fact that only $K_{PAI}^{3D}$ shows signs of saturation for this range of probe intensities. Circular polarization was used in both trapping configurations.}%
\label{fig:kinten}%
\end{figure}

Consequently, in a second set of experiments, determination of $K_{PAI}$ as a function of intensity was carried out in both trapping configurations (fig.~\ref{fig:kinten}), in order to dig further into dimensionality effects on PAI. The MOT is constantly on, as well as the probe laser, whose polarization is now kept circular.

Rate constant $K_{PAI}$ can actually be seen as an average of the monoenergetic rate constant $K(E,\omega)$ over the distribution of collision energies $f(E,T)$ available in the trap at temperature T :
\begin{equation}
K_{PAI}(T,\omega)=\int_0^\infty{f(E,T)K(E,\omega)dE},
\label{eq:kpaiavarage}
\end{equation}
and if $K(E,\omega)$ is a  priori independent of the geometry of the system, $f(E,T)$ is not. 
Here $\omega$ represents the frequency of light employed, in our case as the second step of PAI. For our experiments this frequency is kept constant.

Varying the intensity of the probe beam and measuring the ion production rate, the interaction volume and the number of atoms, we could calculate $K_{PAI}$ for both an atomotron and a 3D MOT, with a probe laser detuning frequency of 2.7~GHz with respect to the $F=2-F'=3$ transition. The comparison data is shown in fig.~\ref{fig:kinten}. There one can observe that $K_{PAI}^{1D}$ is more than one order of magnitude smaller than $K_{PAI}^{3D}$. Similar behavior was observed for photoassociation in a rubidium atomotron~\cite{Marcassa_2005}. The first contribution to such a ratio is certainly the polarization dependence of $K_{PAI}^{1D}$ that we formerly pointed out. For a circular polarization the observed value $K_{PAI}^{1D}$ is a factor 3.4 lower than the one measured in the 1D case employing a linear polarization parallel to the collision axis. Yet, even after correction for polarization, a factor of approximately 15 is still separating the 1D from the 3D situation.

One hypothesis to explain the remaining difference in the photoassociative ionization rate can be associated with a reduced available phase space in the case of 1D collisions, translated by the fact that the collision energy distribution $f(E,T)$ depends on the dimensionality of the trapped cloud as pointed out in~\cite{RamirezSerrano_2004}. Indeed,
\begin{equation}
f_{1D}(E,T)= \frac{1}{\sqrt{\pi}}\frac{e^{-E/k_BT}}{\sqrt{E/k_BT}}=\frac{k_BT}{2E}f_{3D}(E,T).
\label{eq:f1d}
\end{equation}
From these equations, one can realize that even if $f_{1D}(E,T)$ tends to infinity when $E$ goes to zero, its value remains below the value of $f_{3D}(E,T)$ all over the range of energies concerned by PAI. Indeed, the laser producing the first excitation of the PAI process (repumping laser) is detuned of $\Delta_1=-2\Gamma$ with respect to the $F=1-F'=2$ transition and is preferably resonant with atoms of energy $E_1 = 2\hbar\Gamma\ \approx k_B T$. Therefore, for the concerned range of energies, the distribution function weighs $K(E,\omega)$ more heavily in the 3D case than in the 1D case, leading to higher values of $K_{PAI}$ in the MOT.

However, concerning the overall behavior of PAI as a function of intensity, there is a strong indicative that the curvatures of the graphics are opposite for the MOT and the Atomotron. While the first four points indicate a positive curvature for the Atomotron the opposite is observed with the MOT. This behavior together with the remaining points indicates the occurrence of a possible saturation for the MOT, which seems to be not present in the case of Atomotron. This is however just an hypothesis. Saturation of PAI with intensity was already observed in previous experiments~\cite{Bagnato_1993b}, but only in the case of single color PAI in a MOT. In that system, a local
equilibrium theory, where the atoms interact with the light fields near the Condon points, could not fully explain the results. A more elaborated theory, based on Liouville evolution equation for the density matrix elements, matched better
with the experiment. That approximation was called Optical-Bloch-Equations (OBE) and was developed by Band and Julienne~\cite{Band_1992}.

However, there  is no report on investigation of intensity dependence of two color PAI, and the reason why we observe saturation in the 3D case is not fully understood yet. In the present experiment, the evolution of the PAI rate constant is observed as a function only of the intensity of the second optical excitation of the process. The laser responsible for the first step of PAI (repumping laser) is kept at constant intensity, consequently promoting a constant flux of atomic pairs to the intermediate state $Na + Na^*$. After motion along the attractive branch of this state, the flux of pairs can be intercepted by the probe laser (of adjustable intensity) and promoted to the doubly excited state $Na^* + Na^*$, which can finally evolve to short range ionization. In this situation, saturation can be due to several different factors. According to us, the most important of them would be the limitation of the incoming flux in the intermediate state and the light shift caused by the probe laser.

Since the first PAI step takes place at constant intensity, the excitation rate to the intermediate state $Na + Na^*$ is fixed. Excitation to the doubly excited state and ionization will therefore always be limited by this rate, which can eventually translate into a saturation behavior as the intensity of the second laser is increased. The absence of saturation in the ring configuration could then be due to the constant renewal of the atomic population available for PAI due to the fact that atoms are permanently circulating and that we are only exciting a restricted zone of the cloud. In the 3D case though, all the available pairs are exposed to the PAI laser. Thus after they are accelerated and ionized, production of new pairs in the right configuration for PAI may require a delay due to thermalization or trapping of new atoms, that could lead to saturation of the process.
This is in fact in agreement with the fact that the interaction time is on the order of $10^{-8}$ s while the cooling time is at least 10 times longer. Therefore ionized atoms are not replaced by the cooling process, creating a saturation behavior.

On another hand, as the intensity of the probe increases, the light shifts caused onto the intermediate state may decrease the transfer of population to the doubly excited state and reduce the efficiency of the process, also causing a saturation type effect. However, light shifts should occur regardless of the collision dimensionality (1D or 3D) and this is why we think that saturation of the 3D PAI rate constant with the probe intensity is most probably due to the limitation of pairs entering the intermediate channel $Na + Na^*$. But the variations on the wave-packet dynamics in both dimensionalities may be such that a difference is expected for the investigated range of intensities. Nevertheless, full explanation provided with a OBE
type model is clearly out of the scope of this report, though we hope that some interest may rise for the complex theoretical aspects of this problem. In any case, this is just an hypothesis which has to be further experimentally investigated.


Both experiments confirm that collisional behaviors can be very distinct, depending on the trap geometry. The fact that the collision axis is determined restricts the available collision space and modifies the rate of the process as well as its dependence with light intensity and polarization. Besides opening a door to the relatively unexplored world of 1D chemistry, present results may find relevance on modern schemes for quantum computation where 1D arrays of neutral atoms are proposed~\cite{Julienne}. 
The implementation of such schemes will require understanding of the possible modification of collisional and shielding effects~\cite{Gorshkov_2008} in 1D, and gives motivation for further investigation of cold collisions in low dimensions.

{\small The team acknowledges Daniel V. Magalhães for fruitful discussion and technical support. We would like to thank the financial support from FAPESP, CNPq and CAPES.}


{\small
\begin{thebibliography}{}
\bibitem{Weiner_1999} J. Weiner \textit{et al.}, Rev. Mod. Phys. \textbf{71}, 1 (1999).
\bibitem{Degraffenreid_2000} W. DeGraffenreid \textit{et al.}, Rev. Sci. Instr. \textbf{71}, 3668 (2000) and ref. therein.
\bibitem{Marcassa_2005} L.G. Marcassa \textit{et al.}, Phys. Rev. A \textbf{72}, 060701(R) (2005).
\bibitem{Weiner_1989} J. Weiner, J. Opt. Soc. Am. B \textbf{6}, 2270 (1989).
\bibitem{Lett_1993} P.D. Lett \textit{et al.}, Phys. Rev. Lett. \textbf{71}, 2200 (1993).
\bibitem{Paiva_2009} R.R. Paiva \textit{et al.}, Las. Phys. Lett. \textbf{6} (2), 163 (2009).
\bibitem{Phillips_1982} W. D. Phillips and H. Metcalf, Phys. Rev. Lett. \textbf{48}, 596 (1982).
\bibitem{Bagnato_1993} V.S. Bagnato \textit{et al.}, Phys. Rev. A \textbf{48}, 3771 (1993).
\bibitem{Dias_1996} F. Dias Nunes \textit{et al.}, Phys. Rev. A \textbf{54}, 2271 (1996).
\bibitem{Gallagher_1991} A. Gallagher, Phys. Rev. A \textbf{44} (7), 4249 (1991).
\bibitem{Stwalley_1978} W.C. Stwalley \textit{et al.}, Phys. Rev. Lett. \textbf{41}, 1164 (1978).
\bibitem{RamirezSerrano_2002} J. Ramirez-Serrano \textit{et al.}, Phys. Rev. A \textbf{65}, 052719 (2002).
\bibitem{RamirezSerrano_2004} J. Ramirez-Serrano \textit{et al.}, Phys. Rev. A \textbf{69}, 042708 (2004).
\bibitem{Araujo_1995} M.T. de Araujo \textit{et al.}, Opt. Comm. \textbf{119}, 85 (1996).
\bibitem{Bagnato_1993b} V.S. Bagnato \textit{et al.}, Phys. Rev. A \textbf{48}, R2523 (1993).
\bibitem{Band_1992} Y.B. Band and P.S. Julienne, Phys. Rev. A \textbf{46}, 330 (1992).
\bibitem{Julienne} P.S. Julienne, private communication.
\bibitem{Gorshkov_2008} A.V. Gorshkov \textit{et al.}, Phys. Rev. Lett. \textbf{101}, 073201 (2008).
\end{thebibliography}
}
\end{document}